\documentstyle[12pt]{article}
\begin{document}
\setlength{\baselineskip}{0.27in}

\newcommand{\beq}{\begin{equation}}
\newcommand{\eeq}{\end{equation}}
\newcommand{\beqa}{\begin{eqnarray}}
\newcommand{\eeqa}{\end{eqnarray}}
\newcommand{\lsim}{\begin{array}{c}\,\sim\vspace{-21pt}\\< \end{array}}
\newcommand{\gsim}{\begin{array}{c}\sim\vspace{-21pt}\\> \end{array}}
\newcommand{\nt}{\nu_\tau}
\newcommand{\nee}{\nu_e}
\newcommand{\nm}{\nu_\mu}
\newcommand{\bohr}{\mu_B}

\begin{titlepage}
{\hbox to\hsize{Sept 1993\hfill } }
{\hbox to\hsize{Bartol-930XXX\hfill JHU-TIPAC-930026}}
{\hbox to\hsize{UM-TH-93-26\hfill hep-ph/9310349}}
\begin{center}
\vglue .06in
{\Large \bf Closing the Windows on MeV Tau Neutrinos} \\[.5in]

\begin{tabular}{c}
{\bf K. S. Babu\footnotemark[1]}\\[.05in]
{\it Bartol Research Institute}\\
{\it University of Delaware}\\
{\it Newark DE 19716}\\[.15in]
\end{tabular}
\begin{tabular}{c}
{\bf Thomas M. Gould\footnotemark[2]}\\[.05in]
{\it Department of Physics and Astronomy}\\
{\it The Johns Hopkins University}\\
{\it Baltimore MD 21218 }\\[.15in]
\end{tabular} \\
\begin{tabular}{c}
{\bf I.Z. Rothstein\footnotemark[3]}\\[.05in]
{\it The Randall Laboratory of Physics}\\
{\it University of Michigan}\\
{\it Ann Arbor MI 48109 }\\[.15in]
\end{tabular}
\vskip 0.25cm
\footnotetext[1]{babu@bartol.udel.edu}
\footnotetext[2]{gould@fermi.pha.jhu.edu}
\footnotetext[3]{irothstein@umiphys.bitnet}

{\bf Abstract}\\[-0.05in]

\begin{quote}

In this note, we analyze various constraints on the ``visible''
decay modes of a massive $\tau $ neutrino,
$\nu_\tau\rightarrow\nu^\prime\,\gamma $
and $\nu_\tau\rightarrow\nu^\prime\, e^+ e^-$,
where $\nu^\prime$ is a light neutrino.  The BEBC beam dump
experiment provides model-independent constraints on these modes.
The lifetime for the $\nu^\prime\, e^+e^-$ mode is constrained to
be $\tau_{\nu^\prime\, e^+e^-} \ge 0.18~(m_{\nu_\tau}/MeV)~sec.$
We point out that the same experiment implies a similar constraint
on the $\nu^\prime\,\gamma$ mode.  This results in a new upper
limit on the transition magnetic moment of $\nu_\tau$,
$\mu_{\rm tran} \le 1.1 \times 10^{-9} (MeV/m_{\nu_\tau})^2 \mu_B $.
Furthermore, a limit on the electric charge of $\nu_\tau$ may be
obtained, $Q_{\nu_\tau} \le 4 \times 10^{-4}e$.  Combining these
constraints with those arising from supernova observations and
primordial nucleosynthesis calculations, we show that these ``visible''
decays cannot be the dominant decay modes of the $\tau$ neutrino.

\end{quote}
\end{center}
\end{titlepage}
\newpage

The tau neutrino ($\nt$) has eluded direct detection thus far.
Yet, its existence has been inferred,
along with some salient properties such as its spin,
from $\tau$ lepton decay and neutrino interaction data.
At present, there is an  experimental bound on its mass,
$m_{\nt} \le 31~ MeV$~\cite{MASS}.

A $\nt$ with a mass in the MeV range has profound
implications for laboratory experiments,
as well as for cosmology and astrophysics.
For example,
it has recently been pointed out  that
$m_{\nt}\ge 0.3~ MeV$ (for Dirac neutrino) or
$\ge 0.5~ MeV$ (for Majorana neutrino) will
contradict primordial nucleosynthesis calculations,
if the $\nt$ lifetime is longer than ${\cal O}(100)$ seconds~\cite{DOL}.
This constraint is independent of the decay products,
and is a consequence of the fact that the energy density of
a non-relativistic species decreases with the cosmic scale
factor $R$ as $R^{-3}$,
while that of a massless species decreases as $R^{-4}$.
Thus,
we may conclude that an MeV $\nt$  should decay
into relativistic particles with a lifetime of less than
${\cal O}(100)$ seconds (or annihilate sufficiently fast),
so as to ameliorate the nucleosynthesis bound\footnote{
If $\nu_\tau$ decays into a light neutrino and a Goldstone boson (Majoron),
the decay lifetime is constrained to be
$< 40~sec.$ for $\nt$ mass in the MeV range~\cite{steig}.}.
This might occur if the neutrino has
a large diagonal magnetic moment~\cite{GKAW},
or possibly a non--zero electric charge~\cite{FL},
which would allow rapid annihilation in the early universe
into $e^+e^-$ pairs.
Alternatively,
$\nt$ may decay into $\nu^\prime\,\gamma$ or
into $\nu^\prime e^+e^-$,
where $\nu^\prime$ stands for a lighter neutrino
(ie.- $\nu_e$ , $\nu_\mu$, a sterile neutrino, or their antiparticles).
In this note,
we use the BEBC (Big European Bubble Chamber) beam dump
experiment (WA66 collaboration) to greatly constrain the possible
visible decay modes of $\nt$.
Then, combining these constraints with the nucleosynthesis bound
and constraints arising from supernova observations,
we rule out the ``visible'' modes as the dominant ones.

In Ref.~\cite{CS},
Cooper--Sarkar et al. have shown how the BEBC beam dump experiment
restricts the diagonal magnetic moment of a stable $\nu_\tau$
to be $\mu_{\rm diag}\le 5.4 \times 10^{-7} \mu_B$,
thus severely restricting the cosmological
annihilation scenario~\cite{GKAW}.
This bound is the consequence of a limit on the rate
of $\nt$ scattering into electrons.
It is noted here
that the transition magnetic moment of $\nu_\tau$ is also
constrained by the same experiment.
We determine a new upper limit,
$\mu_{\rm tran} \le 1.1 \times 10^{-9}~(MeV/m_{\nu_\tau})^2~\mu_B $,
from the non-observation of the radiative decay,
$\nu_\tau \rightarrow \nu^\prime\,\gamma$.
The lifetime for the decay $\nt\rightarrow\nu^\prime\, e^+e^-$
is bounded by
$\tau_{\nu^\prime e^+e^-}\gsim 0.18~(m_{\nt}/MeV)~sec.$
Both of these limits hold for arbitrary $\nt$ masses,
but assume that the $\nt$ lifetime is $\tau\gsim 10^{-12}~sec.$
Furthermore,
the electric charge of $\nt$ may be bounded by
$Q_{\nt} \le 4 \times 10^{-4}e$,
which is comparable to the limit from the SLAC beam dump
experiment~\cite{SLAC}.

The BEBC beam dump experiment produces so-called ``prompt'' neutrinos,
from the decays of heavy charmed mesons $D$ and $D_s$,
produced upstream where a proton beam impacts on a fixed
target~\cite{GRA}.
The target is sufficiently thick to re-absorb the lighter
mesons,
$K$ and $\pi$, before their decay, thus suppressing the
production of non-prompt relative to prompt neutrinos.
The experiment thereby offers a wide kinematic window on
neutrino masses, roughly $M_\nu \lsim {\cal O}(M_D)$.
The results in this paper will only require that the $\nt$'s
be produced in the decay of the $D_s$ mesons,
$ D_s\rightarrow\tau\,\nt\,$.
This allows us to constrain tau neutrino masses in the
range,
$m_{\nt} \lsim m_{D_s} - m_\tau\simeq 180\, MeV$,
which greatly exceeds the present experimental bound of
$31\, MeV$~\cite{MASS}.
The beam dump experiment has been used in the past to obtain
stringent bounds on production and decays of $\nt$.
The results were presented as limits on the mixing angles
in the leptonic sector~\cite{BEBC}.

If the $\nt$'s are sufficiently long-lived,
they will bypass the bubble chamber before decaying.
Conversely, if they are sufficiently short-lived,
they will all decay before reaching the bubble chamber
and again no decays will be observed.
Thus,
a null result in the search for $\nt$ decays
implies both an upper and lower bound on the lifetime.
The number of decays expected to be seen in a detector of
length $d~(\sim 1\, m)$ at a distance $L~(\sim 400\, m)$
from the source is given by
\beq
\label{eq:N}
N \: = \: \Phi\left(\nt\right)\,\exp\left\{-L/\gamma\,\tau\right\} \,
\left[\, 1 \, - \, \exp\left\{-d/\gamma\,\tau\right\}
\,\right] \,
\left(\tau/\tau_p\right)\, A \, \epsilon \, ,
\eeq
for a given flux $\Phi\left(\nt\right)$ of tau neutrinos.
The Lorentz factor is denoted by $\gamma$, and the partial
width to the observed channel is $1/\tau_p$.
The acceptance $A$ is determined by the detector geometry with
a Monte Carlo simulation.
The efficiency $\epsilon$ of the detector is determined from
the efficiencies of the various detector elements
for detecting the products in a given decay channel.
Eq.~\ref{eq:N} reduces in the limit of a small total lifetime,
$ d/\gamma\tau \ll L/\gamma\tau \ll 1$, to
\beq
\label{eq:Nexpanded}
N \: \simeq \: \Phi\left(\nt\right)\,d\, A\,\epsilon/\gamma\,\tau_p
\, .
\eeq
Since the $\nu_\tau $ flux in the WA66 experiment was
${\cal O}(10^7)~cm^{-2}$,
the detector volume was $\simeq~16.6~m^3$,
and the average neutrino energy was ${\cal O}(10)~GeV$,
it is clear from eq.~\ref{eq:Nexpanded}  that the experiment
was sensitive to a partial lifetime $\tau_p \sim {\cal O}(1)~ sec.$
for $m_{\nt}\sim {\cal O}(1)~MeV$.
The bounds on the partial lifetime $\tau_p $ to be explained below
will not apply if
\beq
\label{eq:shortlife}
\tau \: \lsim \:
2.5 \times 10^{-12} \,\left\{ {m_{\nt}\over MeV} \right\} sec. \, ,
\eeq
in which case the number of events recorded in the bubble chamber will
be $\lsim 1$.

No events were observed in the experiment consistent with
radiative $\nt\rightarrow\nu^\prime\,\gamma$ decay~\cite{PC}.
This implies a model--independent lower bound on the
partial lifetime~\cite{PC}
\beq
\label{eq:BEBC}
\tau_{\nu^\prime \gamma}
\: \gsim \: 0.15 \left\{m_{\nt}\over MeV\right\} \, sec \, .
\eeq
This constraint leads immediately to an upper bound on the
transition magnetic moment of the tau neutrino, $\mu_{\rm tran}$.
The partial lifetime for the radiative decay, due to
a transition magnetic moment $\mu_{\rm tran}$, is
\beq
\label{eq:magmom}
\tau^{-1}_{\nu^\prime \gamma} \: = \:
{\alpha \over 8} \left({\mu_{\rm tran} \over {\mu_B}}\right)^2
\left({ m_{\nt} \over m_e}\right)^2 \, m_{\nt}
\eeq
resulting in the bound
\beq
\mu_{\rm tran} \:\lsim\: 1.1 \times 10^{-9} \,
\left\{ {MeV\over m_{\nt}} \right\}^2 \,\mu_B\, .
\eeq
This bound is much more stringent for MeV $\nu_\tau$
than the corresponding bound for the diagonal magnetic moment~\cite{CS},
and is valid for arbitrarily small $\nt$ mass.

An MeV $\nt$  may also decay into \mbox{$\nu^\prime e^+ e^-$},
where $\nu^\prime$ is $\nu_e$, $\nu_\mu$, a sterile neutrino,
or their antiparticles.
The CHARM experiment rules out this possibility for $\nt$
masses greater than $10~MeV$ if the decay is rapid~\cite{CHARM}.
For the decay into $\nu^\prime e^+e^- $,
the BEBC experiment provides a model--independent bound similar to
eq.~\ref{eq:BEBC}~\cite{PC}:
\beq
\label{eq:BEBC2}
\tau_{\nu^\prime e^+e^-}
\: \gsim \: 0.18\left\{m_{\nt}\over MeV\right\} \, sec \hspace{0.5cm}
{\rm if} \hspace{0.5cm} m_{\nt}\gsim 2\, m_e \, .
\eeq
The slight ($20$\%) improvement relative to eq.~\ref{eq:BEBC}
is due to the difference in conversion efficiencies.
Note that this limit is model--independent,
and does not assume the decay to occur via neutrino mixing in
the charged current.
As such,
it applies to scenarios where the decay is mediated by
exotic particles.

The BEBC beam dump experiment also implies an upper limit
on the electric charge of $\nu_\tau$, $Q_{\nu_\tau} = q\, e$,
from a consideration of the elastic scattering,
$\nt\, e^- \rightarrow \nt\, e^-$.
(The weak contributions are too small and can be ignored~\cite{BV}.).
The cross section for scattering into a forward cone defined
by the BEBC cut on the electromagnetic shower energy,
$T_e \gsim T_{min} = 0.5~GeV$, is
\beq
\sigma \: \simeq \:  4\pi\, r_e^2 \:
\left(\,{q^2\over 32\pi^2}\,\right)\:
\left(\,{m_e\over T_{min}}\,\right) \, ,
\eeq
in the limit $m_e \lsim m_{\nt} \ll T_{min} \ll E \sim 20~GeV$.
Comparing this to the upper bound on the cross-section
implied by the upper bound on the diagonal magnetic moment~\cite{CS},
we obtain $q \le 4 \times 10^{-4}$.
This bound is comparable to the SLAC beam dump limit~\cite{SLAC}.
It may be possible to strengthen this bound by a detailed Monte Carlo
simulation.

The BEBC beam dump limits are complementary to the various
constraints on MeV $\nt$ from cosmology and astrophysics.
For the most part,
these indirect limits from cosmology and astrophysics are not applicable
when the lifetimes become very short.
Eqs.~\ref{eq:BEBC}, \ref{eq:BEBC2} and \ref{eq:shortlife} constrain
the lifetime of an MeV $\nu_\tau$ to be either greater than a second
or less than about $10^{-12}$ seconds.
We summarize the relevant cosmological and astrophysical limits and
show how both these allowed windows are excluded for dominant decays
into visible modes.

The radiative lifetime can be bounded from gamma ray observations
by the Solar Maximum Mission Satellite which was in operation at
the time when the supernova \mbox{SN1987A} explosion was reported.
In Ref.~\cite{SMM},
it was found that for neutrinos with masses less than about \mbox{$50~MeV$,}
the radiative lifetime must satisfy
\beq
\label{eq:SN}
\tau_{\nu^\prime\gamma}\: > \:
8.4\times 10^8\, \left\{MeV \over m_{\nt}\right\} \, sec \, .
\eeq
A similar bound applies to the lifetime for
$\nt\rightarrow\nu^\prime e^+e^-$ decay~\cite{LUMINOSITY,NUSS}.
However,
these bounds do not apply if the decay of $\nu_\tau$
is so rapid that the photon gets trapped inside the progenitor,
which has a radius of $R_{pro}\lsim 3\times 10^{12}\, cm$.
Using the temperature of the neutrino sphere to be \mbox{$T_{\nu}\sim 6~MeV$,}
we see that the constraint on decay does not apply if
\beq
\label{eq:progenitor}
\tau \: \lsim \:
\left\{
\begin{array}{lll}
10\,\left( {m_{\nt}/MeV}\right) &sec., &
\; {\rm if} \; m_{\nt} \lsim 10~MeV \\
50\,\left( {m_{\nt}/MeV}\right)^{1/2} &sec., &
\; {\rm if} \; m_{\nt} \gsim 10~MeV \nonumber
\end{array}
\right. \, .
\eeq
For shorter lifetimes,
there are other bounds which must be considered.
It has been shown that,
if the neutrino decays within the progenitor into visible channels,
then the energy which it deposits ($10^{53}~{\rm ergs}$)
will greatly enhance the supernova luminosity,
thus conflicting with the measured light curves.
However, if the neutrino decays within the neutrino-sphere,
then its visible decay products will thermalize,
thus avoiding the constraint from the supernova
luminosity~\cite{LUMINOSITY,NUSS}.
Assuming the neutrino-sphere radius is be
${\cal O}(10)\, km\,\sim\, {\cal O}(10^{-6})\, R_{pro}$,
we find that the supernova luminosity (SNL) bound does not apply
if the lifetime is less than ${\cal O}(10^{-6})$ times the
bound in eq.~\ref{eq:progenitor}.
This bound is complementary to the regions ruled out by BEBC,
as seen in Fig. 1.

While it is true that these supernova bounds hold for Dirac
as well as Majorana neutrinos,
it has been noted~\cite{seckel} that the supernova data may
be used to rule out Dirac neutrinos with masses greater than
${\cal O}(20)~keV$.
However, it is possible in some models to evade this bound~\cite{babu}.
For MeV neutrinos,
the right handed species is already in thermal equilibrium.
Therefore, models for trapping are less constrained.
The constraint is  model dependent and will not be considered
further here.

Thus,
we see that the only allowed window for visible decay modes is when
$\nt$ decays so rapidly that its lifetime satisfies
eq.~\ref{eq:shortlife}.
However,
we are able to show that neither $\nu^\prime\,\gamma$ nor
$\nu^\prime\, e^+e^-$ can be the dominant decay mode
satisfying eq.~\ref{eq:shortlife}.
Suppose that the radiative decay  dominates. From the
experimental bound on the $\nu_\tau$ magnetic moment, viz.,
$\mu \le 4 \times 10^{-6}\mu_B$~\cite{grotch},
which holds for both diagonal as well
as transition moments, we first derive a limit
$m_{\nt} \gsim 8~MeV$ for eq.~\ref{eq:shortlife} to be satisfied.
For $m_{\nt}$ in the range $8-31~MeV$,
there is an open window for radiative decay,
if the transition magnetic moment is near the experimental limit.
However, if the decay product involves $\nu_e$, $\nu_\mu$ or
their antiparticles,
we can use the better experimental limits on $\mu_{\rm tran}$
for  $\nu_e$ and $\nu_\mu$.
These limits are
$\mu_{\rm tran} \le 1.08 \times 10^{-9}~\mu_B$ for $\nu_e$,
and
$ \mu_{\rm tran} \le 7.4\times 10^{-10}~\mu_B$ for $\nu_\mu$.
So the decay $\nu_\tau \rightarrow \nu_{e,\mu}\,\gamma$ cannot
satisfy eq.~\ref{eq:shortlife}.
This leaves decay into a sterile species as the only option.
But if a sterile species ($\nu_s$) is involved,
the decay may be constrained from nucleosynthesis~\cite{yazaki},
since $\nu_s$ will contribute to the energy density.
Helium may be overproduced for lighter $\nu_{\tau}$ masses.
Thus, the window for the rapid decay into $\nu_s\,\gamma$ may be
closed further with a detailed nucleosynthesis calculation.

Similar arguments can be used to show that $\nu_\tau \rightarrow
\nu^\prime e^+e^-$ cannot be the dominant decay.
There is an upper limit from the search for
$e^+e^- \rightarrow \nu\,\bar{\nu}\,\gamma$
on the effective four--Fermion interaction
$G_{\rm eff} \lsim 6~G_F$~\cite{asp},
where $G_F$ is the Fermi coupling.
As a result, the lifetime will be $\ge 10^{-4}~sec.$,
outside the range in eq.~\ref{eq:shortlife}.

Of course,
none of the  constraints mentioned above shed any light on
neutrino decay into a light neutrino and scalar~\cite{steig}.
Such decays are indeed predicted in many models where
the see-saw mechanism is used to generate a neutrino mass.
Similarly, invisible decays into three light neutrinos also seem to be a
viable scenario~\cite{nnn}.
The results of this paper seem to suggest that if the $\nu_\tau$ mass
is in the MeV range,
these invisible decays are the only possibilities.

\vspace{0.2cm}
\centerline{\bf Acknowledgements}
The authors have benefitted from numerous communications with
S. Sarkar regarding the BEBC experiment.
They wish to thank A.M. Cooper-Sarkar for running the necessary
Monte Carlos and providing  eq.~\ref{eq:BEBC} and eq.~\ref{eq:BEBC2}.
They also wish to thank S. Nussinov, D. Seckel, G. Steigman
and M. Turner for discussions.
IZR and TMG gratefully acknowledge the hospitality of the Aspen Center
for Physics where some of this work was completed.
TMG acknowledges the support of the National Science
Foundation under grant \mbox{PHY-90-9619} and the Texas National Research
Lab Commission under grant \mbox{RGFY-93-292.}
IZR and KSB are supported in part by grants from the Department of Energy.

\vspace{0.2cm}
\centerline{\bf Figure Caption}
\noindent 1. Bounds on radiative and $e^+e^-$ decays of $\nt$.\\
Curve labels lie on the forbidden side of curves:
lab mass bound (Argus), nucleosynthesis (NS),
supernova luminosity (SNL), Solar Max Mission (SMM).
The bounds apply to the radiative decay for all $m_{\nt}$ shown,
but apply to the $e^+e^-$ decay only for $m_{\nt}\gsim 2~m_e$.
The hatched region for $\tau \lsim 10^{-10} sec.$
is consistent with all bounds plotted, but is ruled out
as the dominant decay mode, for non-sterile neutrinos,
by the analysis on pg.~5.

\end{document}